\documentclass[journal=jpclcd]{achemso}
\setkeys{acs}{maxauthors=10,etalmode=truncate} 
\usepackage{amsmath}
\usepackage{color}
\usepackage{txfonts}
\usepackage{mathrsfs}
\usepackage{dcolumn}
\usepackage{bm}
\usepackage{multirow}
\usepackage{graphicx}
\usepackage{rotating}
\usepackage{layout}
\usepackage[version=3]{mhchem}
\usepackage{braket}
\usepackage{longtable}
\usepackage{setspace}

\allowdisplaybreaks
\SectionNumbersOn
\usepackage{xfrac}

\usepackage[colorlinks,citecolor=blue,urlcolor=blue,bookmarks=false,hypertexnames=true]{hyperref} 

\title{Exploring Ligand-to-Metal Charge-transfer States in the Photo-Ferrioxalate System using Excited-State Specific Optimization}
\date{\today}
\author{Lan Nguyen Tran}
\email{tnlan@hcmiu.edu.vn}
\affiliation{Department of Physics, International University, Ho Chi Minh City, Vietnam}
\alsoaffiliation{Vietnam National University, Ho Chi Minh City, Vietnam}
\author{Eric Neuscamman}
\email{eneuscamman@berkeley.edu}
\affiliation{Department of Chemistry, University of California, Berkeley, California, 94720, USA}
\alsoaffiliation{Chemical Sciences Division, Lawrence Berkeley National Laboratory, Berkeley, CA, 94720, USA}

\begin{document}

\begin{abstract}
The photo-ferrioxalate system (PFS), [Fe(III)(C$_2$O$_4$)]$^{3-}$, more than an exact chemical actinometer, has been extensively applied in wastewater and environment treatment. Despite many experimental efforts to improve clarity, important aspects of the mechanism of ferrioxalate photolysis are still under debate. In this paper, we employ the recently developed W$\Gamma$-CASSCF to investigate the ligand-to-metal charge-transfer states key to the ferrioxalate photolysis. This investigation
provides a qualitative picture of these states and key potential energy surface features
related to the photolysis. Our theoretical results are consistent with the prompt charge transfer
picture seen in recent experiments and clarify some features that are not visible in experiments. Two ligand-to-metal charge-transfer states contribute to the photolysis of ferrioxalate, and the avoided crossing barrier
between them is low compared to the initial photoexcitation energy. Our data also clarify
that one Fe–O bond cleaves first, followed by the C–C bond and the other Fe–O bond.

\begin{tocentry}
\includegraphics[width=5.0cm]{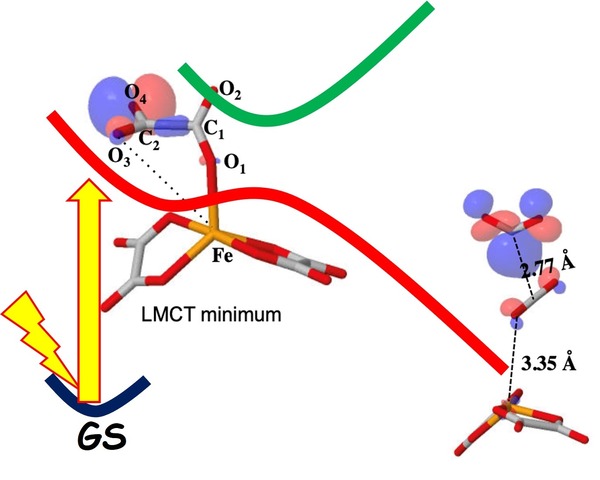}
\end{tocentry}

\end{abstract}

\maketitle

Photo-ferrioxalate system (PFS) [Fe(III)(C$_2$O$_4$)]$^{3-}$ has been used innumerable times for the chemical actinometry in liquid systems\cite{PFS_act1,PFS_act2,PFS_act3,PFS_act4,PFS_act5}. Recently, it has been extensively applied in wastewater and environment treatment\cite{PFS_app1,PFS_app2,PFS_app3,PFS_app4}. The photolysis of PFS generates various reactive intermediates, such as Fe(II) and reactive oxygen species (ROS), including O$^{\bullet-}$, $^{\bullet}$OH and H$_2$O$_2$, that are significant for the oxidative decontamination of organic substances in the environment and engineered systems. This photoreduction is conventionally expressed as
\begin{align}
2[\text{Fe(III)(C}_2\text{O}_4)_3]^{3-} + h\nu &\rightarrow 2[\text{Fe(II)(C}_2\text{O}_4)_2]^{2-} + 2\text{CO}_2 + (\text{C}_2\text{O}_4)^{2-}.       
\end{align}
Understanding its prime mechanisms is crucial to optimize the reactive paths and to control their outcomes. However, while the photoreaction intermediates and products are known, the initial charge-transfer mechanisms that play a pivotal role in the photochemistry of PFS are still under debate\cite{PFS_review2023}. 

Two mechanisms have been proposed for the photolysis of PFS\cite{chen2007electron,chen2008electron,pozdnyakov2008new,chen2009comment,pozdnyakov2009reply}: (i) a prompt reduction and (ii) a slow reduction. In the first mechanism, a ligand-to-metal charge-transfer (LMCT) state is promptly excited, resulting in the reduction of the metal center from Fe(III) to Fe(II)\cite{pozdnyakov2008new,pozdnyakov2009reply}.
The second mechanism has no intramolecular electron transfer, with the metal center
instead reduced later on by an electron transfer from the environment.
\cite{chen2007electron,chen2008electron}
While numerous experiments support the prompt reduction mechanism, specific interpretations of
its nature differ. 
A more comprehensive review of these experimental results is given in the Supporting Information (SI), and we only summarize some important conclusions here.
Pozdnyakov et al. \cite{pozdnyakov2008new} proposed that electron transfer from oxalate to iron
forms an oxalate radical that initially remains bound to the metal and then diffuses into a solution.
Ogi et al. \cite{ogi2015ultraviolet} proposed that electron transfer from oxalate to iron
produces a bound excited molecule that dissociates as CO$_2^{\bullet-}$ and CO$_2$. 
The conclusions of Mangiante et al. \cite{mangiante2017mechanism} and O’Neil et al.
\cite{o2017ultrafast} mostly agree with those of Ogi et al.
Pilz and coworkers \cite{pilz2019time}  showed evidence that, after the LMCT excitation,
the excited molecules dissociate into a single CO$_2$ and the complex
[CO$_2$Fe(II)(C$_2$O$_4$)$_2$]$^{3-}$,
which has been further supported by work by Longetti et al. \cite{longetti2021ultrafast}. 

Although the experimental studies above have provided significant evidence for a prompt LMCT
photoexcitation followed by the dissociation of the oxalate anion, the nature of the initial
charge transfer (CT) and its immediate effects on the complex is still under debate.
\cite{longetti2021ultrafast}
It remains unclear if the species remains bound to the metal center after the
photoinduced CT or whether it immediately dissociates from the metal to
diffuse into the solution. \cite{pilz2019time,longetti2021ultrafast}
Certainly, a detailed ordering of the Fe--O and C--C bond dissociations and any
related transition states on the LMCT surface are also lacking. 

Regarding theoretical studies, density functional theory (DFT) and time-dependent DFT (TD-DFT)
have been used to investigate the LMCT state of PFS
\cite{chen2007electron, chen2008electron,straub2018iron,straub2018femtosecond}. While DFT is only able to optimize the lowest state of a given spin multiplicity,
TD-DFT faces well-known challenges when modeling charge-transfer excitations. 
\cite{dreuw2005single,kummel2017charge}  
While the complete active space self-consistent field (CASSCF) method with
state-averaging optimization has been widely used for excited states. It also
has serious shortcomings for CT states whose important orbital
relaxations and dipole moment changes are resisted by the other states in
the state average.
\cite{domingo2012metal,pineda2018excited,tran2019tracking,tran2020improving}
A theoretical tool more robust to the challenges of LMCT is thus highly desirable
in analyzing the photolysis of PFS. 

In the present study, we employ a recently-introduced excited-state specific
approach \cite{tran2019tracking,tran2020improving} to CASSCF to investigate
PFS's LMCT states.
This method targets the state of interest during iterative CASSCF optimization
according to its physical characteristics, such as its energy and density matrices.
The approach is similar in spirit to the maximum overlap method (MOM) developed
for Hartree-Fock and DFT. \cite{MOM_2008,MOM_2021}
While its current form does not make the approach amenable to simulating the
explicit dynamics of the ferrioxalate photolysis, it can provide a qualitative
picture of the key features of the LMCT potential energy surface (PES)
that explain in clearer detail how an immediate post-excitation Fe--O bond
dissociation is then followed by a low-barrier avoided crossing into
a dissociative state in which the C--C bond and then the other Fe--O bonds
rupture.
These findings are consistent with the experimentally-supported prompt LMCT
picture while also revealing features that are not visible in experiments.

Combining experiment and theoretical simulation, Ogi and coworkers \cite{ogi2015ultraviolet} attributed the LMCT state after the photoexcitation to a transition from an oxalate $\sigma-$bonding to a singly-occupied Fe 3$d$ orbital (see Figure 9 in Ref.~\citenum{ogi2015ultraviolet}). In the present study, we start with a (7e,6o) active space including the same ligand orbital as Ogi {\it et al.}\ and five Fe 3$d$ orbitals (see Figure S1 of the SI).  As we will discuss, post-excitation
orbital relaxations lead to a modified character for this ligand orbital, which
in turn changes the optimal excited state geometry.
We have verified that the same holds in larger
active spaces that we discuss below,
but for simplicity's sake we will begin with
this (7e,6o) minimal active space.

\begin{figure}[h!]
  \includegraphics[width=8cm,]{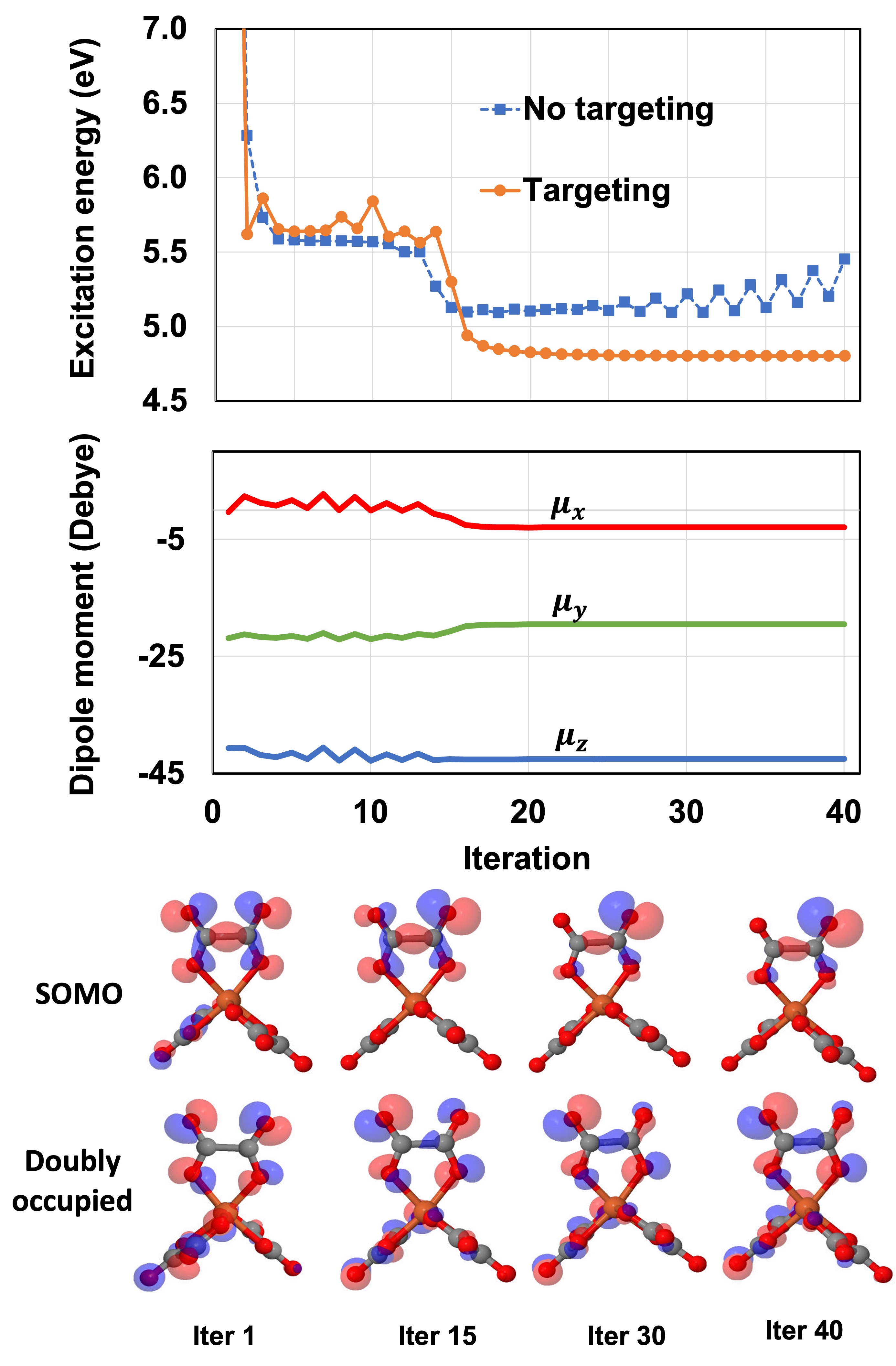}
  \caption{Upper panel: the excitation energy (relative the GS energy, $-64956.96$ eV) of the first LMCT state during SS-CASSCF optimization with and without state targeting using $W\Gamma$ criterion. Midle panel: the change of dipole moment, $\vec{\mu}=\{\mu_x, \mu_y, \mu_z\}$, of the LMCT state during targeted SS-CASSCF. Lower panel: Two ligand's orbitals  of LMCT state at different stages of SS-CASSCF optimization.}
  \label{fig:iter-equi}
\end{figure}

Let us first consider the LMCT state at the ground state (GS) equilibrium geometry. In the top panel of Figure~\ref{fig:iter-equi}, we plot the change of LMCT energy during SS-CASSCF(7e,6o) optimization with and without $W\Gamma$ selection. While SS-CASSCF without targeting exhibits a strong oscillation after several ten iterations and cannot converge, SS-CASSCF with the $W\Gamma$ selection converges to the excitation energy of 4.80 eV, which is quite consistent with the excitation wavelength ($\sim$266/267 nm) observed in experiment \cite{pozdnyakov2008new}. To check whether $W\Gamma$-CASSCF converges to the right state, we plot the change in the dipole moment of the targeted state during optimization in the middle panel. All three dipole moment components do not change significantly, indicating that the converged state is the targeted state. As in the bottom panel, we can see that the shape of the ligand's singly-occupied molecular orbital (SOMO) that loses an electron to an iron $3d$ orbital gradually changes during the SS-CASSCF iteration. While it spreads over the oxalate for the GS, it is dominated by the oxygen lone-pair orbital for the LMCT state. 
Because of the significant difference between the orbital shape of the ground and LMCT states, SA-CASSCF may be inadequate.  

\begin{figure}[t!]
  \includegraphics[width=8cm]{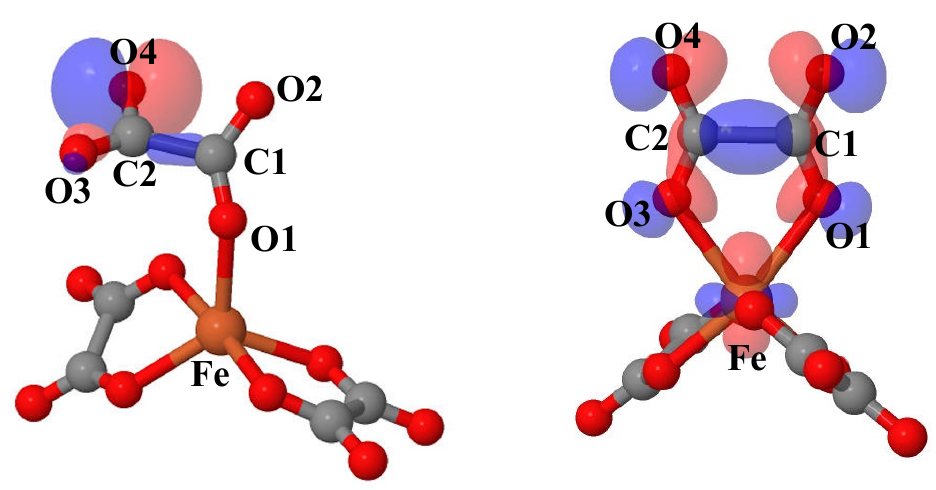}
  \caption{Geometries of b-LMCT equilibrium optimized using $W\Gamma$-CASSCF(7e,6o) (left) and TD-DFT (right). See Table~\ref{tab:b-LMCT} for important parameters. The shapes of $W\Gamma-$CASSCF oxalate's SOMO and TD-DFT oxalate's natural orbital are also presented.}
  \label{fig:ct-equi}
\end{figure}

The question is whether the oxalate anion completely dissociates from the Fe center or remains bound to it after the photoexcitation. To answer this question, we fully relax the geometry of the LMCT state using $W\Gamma$-CASSCF and TD-DFT. 
Geometries and important parameters are presented in Figure~\ref{fig:ct-equi} and Tabel~\ref{tab:b-LMCT}, and Cartesian coordinates are given in the SI. Both methods predict that the LMCT state has a bound minimum, and so we refer to this excited state as b-LMCT. 
Interestingly, however, the two methods have
qualitatively different minimum geometries,
with TD-DFT showing the C1--C2, Fe--O1, and Fe--O3 bonds
all elongated but intact,
and $W\Gamma$-CASSCF showing cleavage of the Fe--O3 bond, slight
elongation of Fe--O1, and essentially no change to C1--C2.
TD-DFT's C1--C2 elongation agrees with the picture presented by
Ogi {\it et al,} \cite{ogi2015ultraviolet} but the reasons behind
these geometric differences, to which we now turn,
casts doubt on the TD-DFT prediction.

To understand why the minimum geometries differ so greatly,
it is crucial to look at both the donor hole orbital (b-LMCT's ligand SOMO)
and the relaxation, or artificial lack thereof, of the other ligand orbitals.
Figure \ref{fig:ct-equi} shows that $W\Gamma$-CASSCF's hole orbital is
localized on an oxygen lone pair, while TD-DFT's is spread across the ligand
in an orbital with substantial $\sigma$ bonding character (hence the
weakened C--C bond).
While localizing the hole may at first glance seem less
stable than spreading it out, the lower panel 
of Figure \ref{fig:iter-equi} reveals that localization is stabilized
by strong relaxations in the shape of a key doubly occupied orbital.
Indeed, even at the GS geometry,
these relaxations allow the hole to localize
without creating undue concentrations of charge on either end
of the ligand.
This localization affects the geometry by creating different
environments for O1 and O3, evidently making it preferable for
O3 to elongate its bond to Fe to the point of cleavage.
In TD-DFT, in contrast, the adiabatic approximation introduces an
artifact in which doubly occupied orbitals do not relax after the excitation. 
\cite{ziegler2009relation,park2015relation}
It thus cannot stabilize a localized hole and instead
places it in the delocalized, partially-$\sigma$-bonding orbital.
As seen in Figure \ref{fig:iter-equi}, $W\Gamma$-CASSCF initially does the same,
but, after state-specific orbital optimization, adopts a relaxation that
stabilizes the localized hole.
Given that TD-DFT's lack of localization seems driven by an artifact of
its approximations, we find the picture painted by $W\Gamma$-CASSCF
to be more persuasive.

We report the vertical and adiabatic excitation energies from $W\Gamma-$CASSCF(7e,6o) and TD-DFT/B3LYP in Table~\ref{tab:energies}. We also consider the effect of solvent on these energy levels using the COSMO solvation model with the water dielectric constant ($\epsilon = 80$). We used geometries optimized without solvent for adiabatic excitation energies with solvent. We can see that TD-DFT excitation energies are lower than $W\Gamma-$CASSCF. Although the solvent generally reduces the excitation energies, its effect on the adiabatic excitation energies of $W\Gamma-$CASSCF is negligible. 

\begin{table}[!h]
  \normalsize
  \caption{\label{tab:b-LMCT} \normalsize Some important parameters of the GS geometry and b-LMCT equilibrium geometries optimized using $W\Gamma-$CASSCF and TD-DFT.}
  \begin{tabular}{ccccccccccccccccccccc}
    \hline \hline		
    \multirow{2}{*}{Parameters} &\multirow{2}{*}{GS} &\multicolumn{3}{c}{b-LMCT}  \\
    \cline{3-5} 
    & &CAS(7e,6o) &CAS(9e,13o) &TD-DFT\\
    \hline
    Fe--O1 (\r{A}) &2.04 &2.21 &2.19 &2.28\\
    Fe--O3 (\r{A}) &2.04 &4.19 &4.10 &2.28\\
    C1--C2 (\r{A}) &1.57 &1.56 &1.58 &1.76\\
    $\alpha$(O1,C1,C2,O3) ($^{\circ}$) &179.83 &46.14 &45.62 &164.50 \\
    $\alpha$(O1,C1,C2,O4) ($^{\circ}$) &0.14 &141.16 &142.54 &15.53 \\
\hline \hline
\end{tabular}
\end{table}

\begin{table}[!h]
  \normalsize
  \caption{\label{tab:energies} \normalsize Energies in eV for $W\Gamma-$CASSCF(7e,6o) and TD-DFT/B3LYP with and without solvent (water). See text for a description
  of the TS barrier calculation.}
  \begin{tabular}{ccccccccccccccccccccc}
    \hline \hline		
    \multirow{2}{*}{Method} \multirow{2}{*}{} &\multirow{2}{*}{GS energy} &\multicolumn{2}{c}{Excitation energy} &\multirow{2}{*}{TS barrier}  \\
    \cline{3-4} 
 & &Vertical &Adiabatic \\
    \hline
 {$W\Gamma-$CASSCF(7e,6o)}&&& \\
    Without solvent &$-$64956.96 &4.80 &2.58 & 0.31\\
    With solvent &$-$64971.02 &3.69 &2.56 &0.64\\
    \hline
TD-DFT/B3LYP&&& \\
    Without solvent &$-$65132.99 &2.55 &0.64\\
    With solvent &$-$65147.99 &2.16 &0.30\\
\hline \hline
\end{tabular}
\end{table}

\begin{figure}[t!]
  \includegraphics[width=12cm]{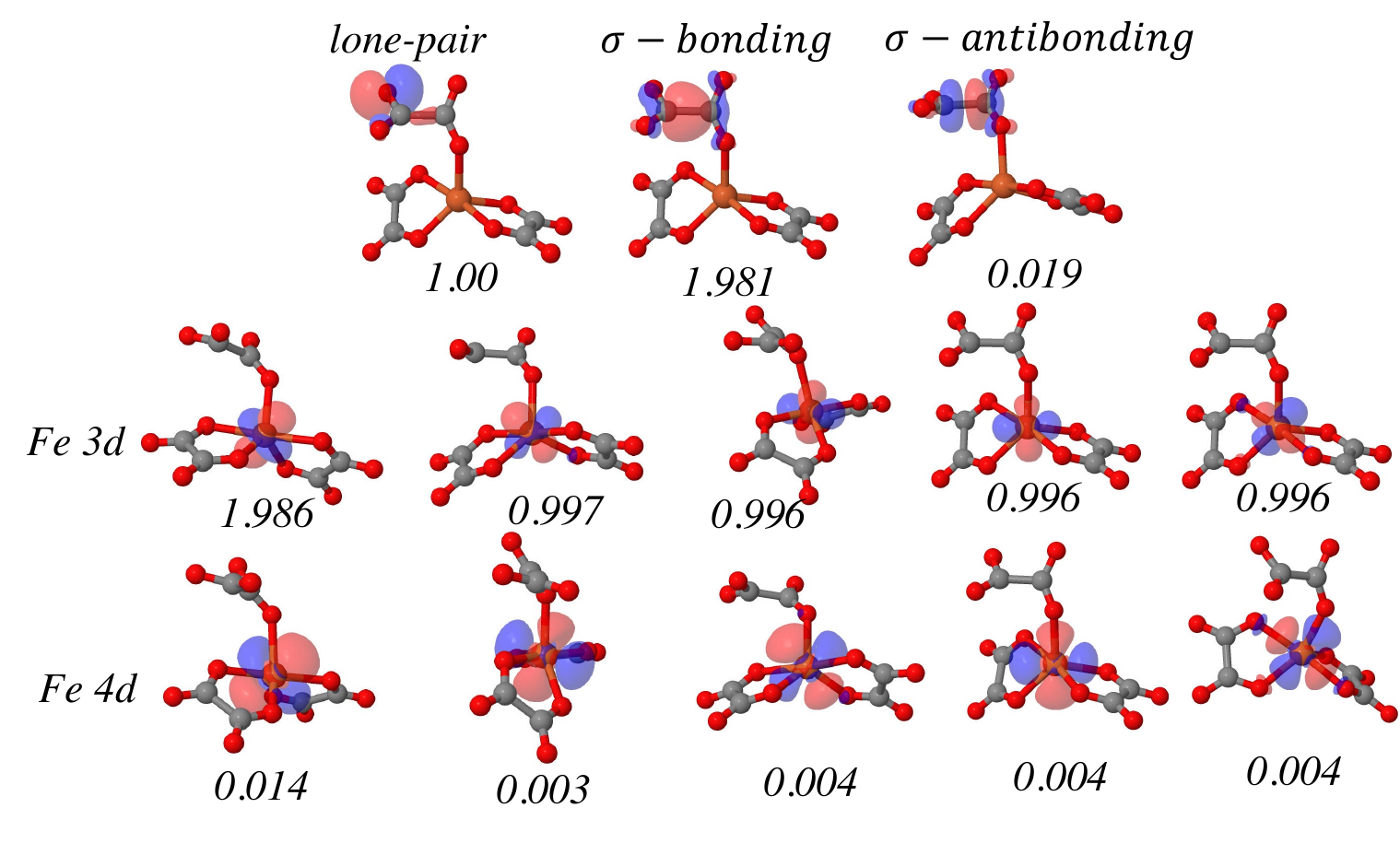}
  \caption{Converged natural orbitals and occupations from the $W\Gamma$-CASSCF(9e,13o) b-LMCT geometry optimization.}
  \label{fig:9e13o_nat}
\end{figure}

To check whether our conclusions are biased by a too-small active
space, we have also employed the expanded (9e,13o) active space shown
in Figure \ref{fig:9e13o_nat}.
This expansion does two things.
First, by including the ligand's $\sigma$ bonding and antibonding orbitals, it
ensures that CASSCF has the flexibility to break the C1-C2 bond if it wants to.
Second, the inclusion of the Fe 4d orbitals guards against the double d shell
effect, in which 4d orbitals can play a larger-than-expected role in transition
metal correlation effects (indeed, Figure \ref{fig:9e13o_nat} shows one of the 4d
having an occupation
similar to the valence antibonding orbital).
We felt this effect was important to be cautious of in this LMCT excitation,
as the two states involved
have differing numbers of d electrons and so will have differing amounts of
d electron correlation.
As seen in Figure \ref{fig:9e13o_nat} and Table \ref{tab:b-LMCT},
neither of these extensions made meaningful changes to the b-LMCT state
geometry or the shape of the donor orbital,
giving us additional confidence in our conclusions above.


We note that our finding of the Fe--O bond cleavage due to intra-molecular CT contrasts with some recent works. Combining experimental spectroscopy and DFT calculations, Chen {\it et al.}\cite{chen2007electron, chen2008electron} suggested that the primary reaction is Fe--O cleavage, but \textit{without} intra-molecular CT. Although Ogi {\it et al.}\cite{ogi2015ultraviolet} provided experimental evidence that the intra-molecular LMCT initiates the photolysis of ferrioxalate, those authors proposed a different primary reaction where the C1--C2 bond weakens and subsequently dissociates, which is consistent with our present TD-DFT results depicted in Table~\ref{tab:b-LMCT} and Figure~\ref{fig:ct-equi}. Recently, O'Neil and coworkers \cite{o2017ultrafast} reported an ultrafast time-resolved X-ray absorption spectroscopy measurement that showed evidence of the elongation of the Fe--O bond. However, the change in the Fe--O bond predicted by those authors is only $0.14\pm0.06$\r{A}, which is comparable to our prediction for Fe--O1 but much smaller than our prediction for Fe--O3.

\begin{figure}[t!]
  \includegraphics[width=10cm,]{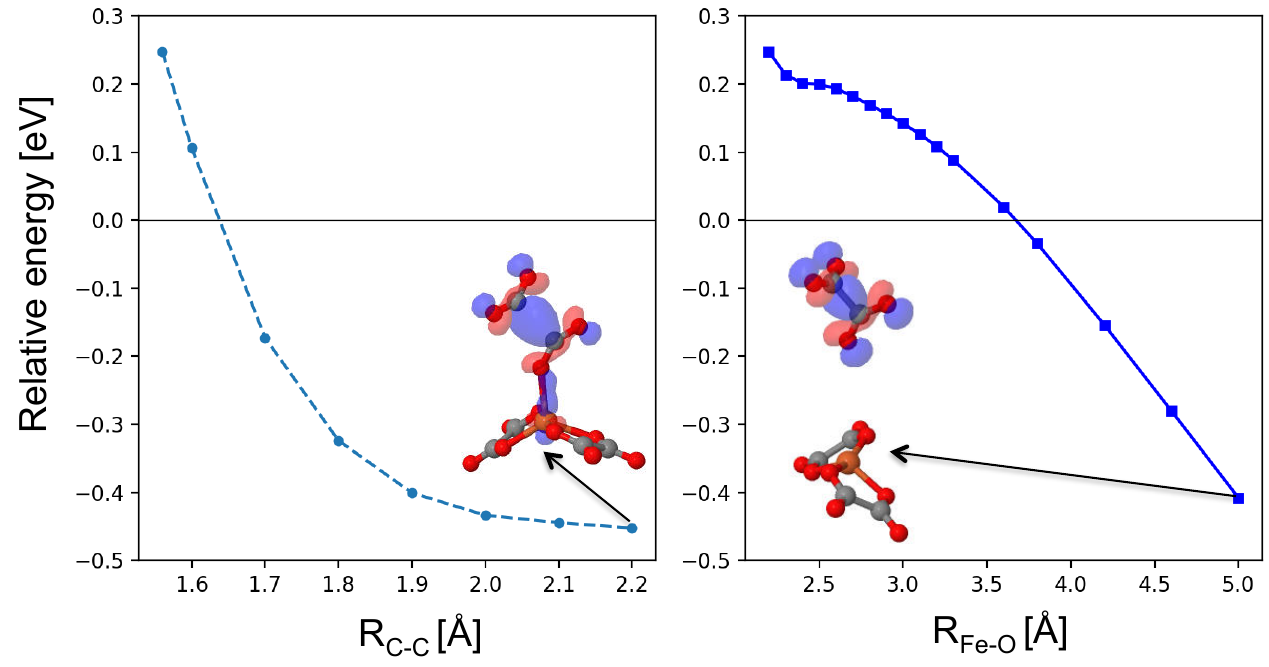}
  \caption{Potential energy curves showing the $W\Gamma-$CASSCF(7e,6o) energy of the d-LMCT state after relaxing the geometry with C1--C2 (left panel) and Fe--O1 (right panel) bond distances held fixed. The Fe--O1 bond is constrained at 2.2\r{A} for the C1--C2 curve, whereas the C1--C2 bond is constrained at 1.56\r{A} for the Fe--O1 curve. The leftmost points of both curves correspond to b-LMCT equilibrium. Ligand's SOMO is dominated by C1--C2 $\sigma-$bonding. Energies are relative to the b-LMCT minimum energy ($-64954.38$ eV), indicated by the horizontal line.}
  \label{fig:ct-stretching}
\end{figure}

The presence of a bound minimum on this b-LMCT state's PES implies that
it has at least a small barrier for the photolysis process,
leading to the question of which bond elongation, Fe--O1 or C1--C2 (see Figure~\ref{fig:ct-equi} for numbering),
is more likely to occur first during the photolysis process.
If, upon Fe--O1 or C1--C2 stretching, a higher state with dissociative
character mixed with the b-LMCT in an avoided crossing, a low barrier
to dissociation would be possible.
To check whether such a dissociative state is present,
we located higher-lying LMCT states at the b-LMCT minimum geometry and
performed unconstrained $W\Gamma-$CASSCF geometry optimizations on them.
These also employed (7e,6o) active spaces as for b-LMCT,
but each had a different ligand donor orbital.
One such optimization was indeed dissociative.
The dissociative state (d-LMCT) appeared to initially
favor C1--C2 stretching, presumably due to the $\sigma$ character of
its donor orbital (see inset images in Figure \ref{fig:ct-stretching}).
We note that initial C1--C2 elongation is consistent with the recent suggestion of Mangiante {\it et al.}\cite{mangiante2017mechanism}, who combined pump/probe mid-infrared transient absorption spectroscopy and theoretical simulations of a single C$_2$O$_4^{\bullet-}$ dissociation (see Figure~5 in Ref.~\citenum{mangiante2017mechanism}).
To clarify that d-LMCT indeed prefers initial C1--C2 stretching instead of
Fe-O1 stretching,
we plot in Figure~\ref{fig:ct-stretching} potential energy curves
during geometry optimizations of this state in which either the C1--C2 or Fe--O1
bond distance is held fixed.
Specifically, we constrained one optimization to have $d_\text{Fe--O1} = 2.21$\r{A}
and the other to have $d_{\text{C1--C2}} = 1.56$\r{A}.
The d-LMCT state showed dissociative character under both of these constraints,
but the rate at which bond elongation lowered the energy during the initial bond
stretching was much greater when stretching C1--C2 than when stretching
Fe--O1, suggesting that a molecule placed on the d-LMCT surface at the
b-LMCT minimum would see its C1--C2 bond elongate first.
This picture
is consistent with recent experiments
\cite{longetti2021ultrafast,straub2018femtosecond,straub2018iron,pilz2019time}
showing that a CO$_2$ molecule is released first, leaving behind
[Fe(ox)$_2$(CO$_2$)]$^{3-}$. 
However, the photochemical process initiated by excitation to b-LMCT
does not start in the dissociative d-LMCT state, so we must now turn
to a more careful analysis of how easy it is for the state character
to switch.

\begin{figure}[h!]
  \includegraphics[width=12cm,]{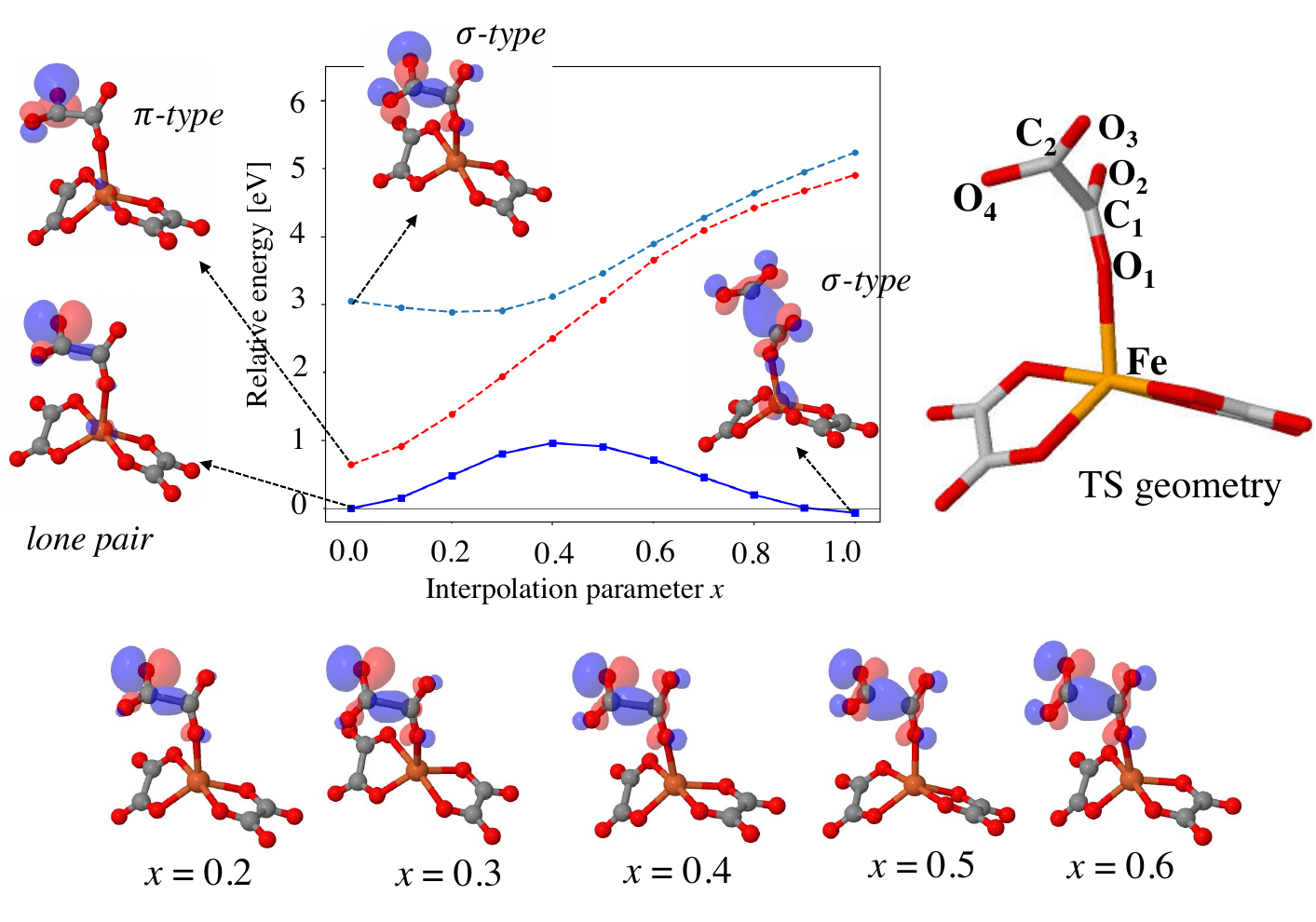}
  \caption{Top left panel: SA(3)-CASSCF(11e,8o) state energies during a
  geometry interpolation between the b-LMCT minimum ($x=0.0$)
  and the minimum geometry for the d-LMCT state when the C1--C2 distance
   is constrained to $R_{\mbox\small{CC}} = 2.2$ \r{A} ($x=1.0$). Three states include b-LMCT, d-LMCT states, and one LMCT state between the two. Energies are relative to the lowest-state energy of SA(3)-CASSCF(11e,8o) at the b-LMCT equilibrium ($-64954.22$ eV). The donor ligand's natural orbitals for different states are also depicted. 
   Top right panel: transition-state (TS) geometry optimized using SA(3)-CASSCF(11e,8o). Bottom panel: ligand's SOMO of the lowest state in SA(3)-CASSCF (the blue solid curve) at different interpolation points $x$.}
  \label{fig:cc-inter}
\end{figure}

\begin{table}[!h]
  \normalsize
  \caption{\label{tab:TS} \normalsize Important geometrical parameters of transition-state (TS) geometry in comparison to the b-LMCT minimum.}
  \begin{tabular}{ccccccccccccccccccccc}
    \hline \hline		
    Parameters &b-LMCT  &TS\\
    \hline
    Fe--O1 (\r{A}) &2.21 &2.32\\
    C1--C2 (\r{A}) &1.56 &1.68 \\
    $\alpha$(O1,C1,C2,O3) ($^{\circ}$) &46.14 &99.9 \\
    $\alpha$(O1,C1,C2,O4) ($^{\circ}$) &141.16 &82.43 \\
\hline \hline
\end{tabular}
\end{table}

If b-LMCT were to smoothly convert into d-LMCT through
an avoided crossing, the energy barrier associated with this crossing
would determine how accessible this this pathway was.
To determine if such an avoided crossing exists at all and to get a good
guess for a corresponding transition state (TS) geometry,
we have performed SA-CASSCF calculations along a linear
interpolation between the Cartesian geometries of the
b-LMCT minimum and the 2.2\r{A} point
from the left panel of Figure \ref{fig:ct-stretching} (see SI for
explicit coordinates for these two geometries).
Since we found that there is in fact another LMCT state in between
b-LMCT and d-LMCT, we have included all three of these LMCT states
in the SA in equal weights.
As each state corresponds to excitation from a different ligand donor orbital
(one lone pair, one $\sigma$, and one $\pi$ as shown in the insets
to Figure \ref{fig:cc-inter}), we employ an (11e,8o) active space
containing the Fe 3d orbitals and these three ligand orbitals
(see Figure S2 in the SI).
As seen in Figure \ref{fig:cc-inter}, the lower adiabat in this interpolation
does indeed show avoided crossing behavior between b-LMCT and d-LMCT:
its donor orbital has clear lone pair
character at the b-LMCT geometry (x = 0), which changes smoothly into
clear $\sigma$ bonding
character on the other side of the barrier (x = 1).

\begin{figure}[t!]
  \includegraphics[width=10cm,]{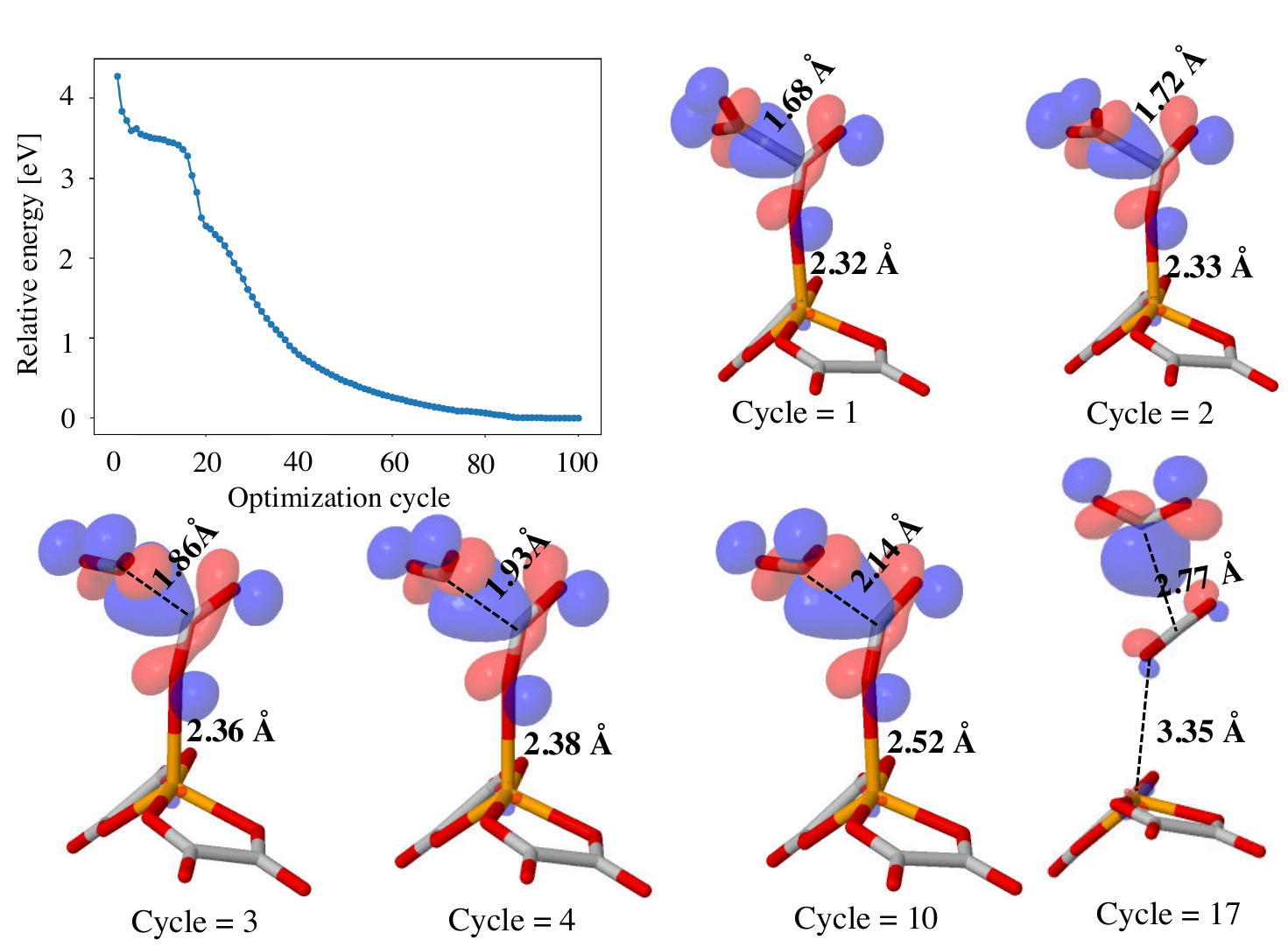}
  \caption{The change of energy, geometry, and ligand's SOMO during the full $W\Gamma-$CASSCF(7e,6o) geometry optimization without constraints starting from the TS geometry (cycle = 1). The energy is relative to the energy at the cycle 100 ($-64958.60$ eV).}
  \label{fig:full-relax}
\end{figure}

Using the geometry at the avoided crossing in the interpolation as an initial guess, we searched for TS geometry using SA(3)-CASSCF(8e,11o). The TS geometry is depicted in the right panel of Figure~\ref{fig:cc-inter}, and some critical geometrical parameters are summarized in Table~\ref{tab:TS}. While both C1--C2 and Fe--O1 bonds elongated slightly compared to the b-LMCT minimum, the oxalate anion is twisted, preparing for the dissociation.
As it is not balanced to compare SA-CASSCF and SS-CASSCF energies,
we then evaluated the $W\Gamma$-CASSCF(7e,6o) energy at the TS and compared it to
the corresponding energy at the b-LMCT minimum.  The results in Table~\ref{tab:energies} show that this TS barrier is small relative to the energy of the initial photoexcitation and that the same is true under implicit solvent.
Such a small barrier relative to the photoexcitation energy suggests that
the CO$_2$ detachment can proceed rapidly following photoexcitation, as has
been measured in recent experiments.
\cite{straub2018femtosecond,pilz2019time,longetti2021ultrafast}

In Figure~\ref{fig:full-relax}, we plot the change of energy, geometry, and
the ligand SOMO during a full relaxation of
the geometry on
$W\Gamma$-CASSCF's representation of the the d-LMCT
surface starting from the
TS geometry. Although both the C1--C2 and Fe--O1 bonds eventually rupture,
one's elongation precedes the other.
For example, in geometry optimization cycle 4, C1--C2 has elongated by
14\%, whereas Fe--O1 has stretched by just 2\%,
again implying that C1--C2 cleavage somewhat precedes Fe--O1 cleavage,
as was suggested by our constrained optimizations
in Figure~\ref{fig:ct-stretching}.
That said, Figure~\ref{fig:full-relax} also shows that, once the
avoided crossing barrier has been crossed, the dissociation of these
two bonds can occur in a single, concerted process. 


We have applied the recently developed $W\Gamma$-CASSCF method to
investigate the LMCT states governing ferrioxalate's photolysis while
fully accounting for post-excitation orbital relaxations, overcoming
a key shortcoming of many existing theoretical approaches.
Many recent experiments have provided evidence that LMCT states play
the pivotal role in the photolysis of ferrioxalate, which makes the
lack of proper excited-state-specific orbital relaxation in
TD-DFT and SA-CASSCF a severe concern.
The present study shows that CASSCF geometry optimization on the LMCT state
is possible with $W\Gamma$-CASSCF despite SA-CASSCF's difficulties in
this regard.
Although we have not simulated the explicit dynamics of
ferrioxalate photolysis, our results do clarify key qualitative
features on the LMCT potential energy surface.

\begin{figure}[t!]
  \includegraphics[width=8cm,]{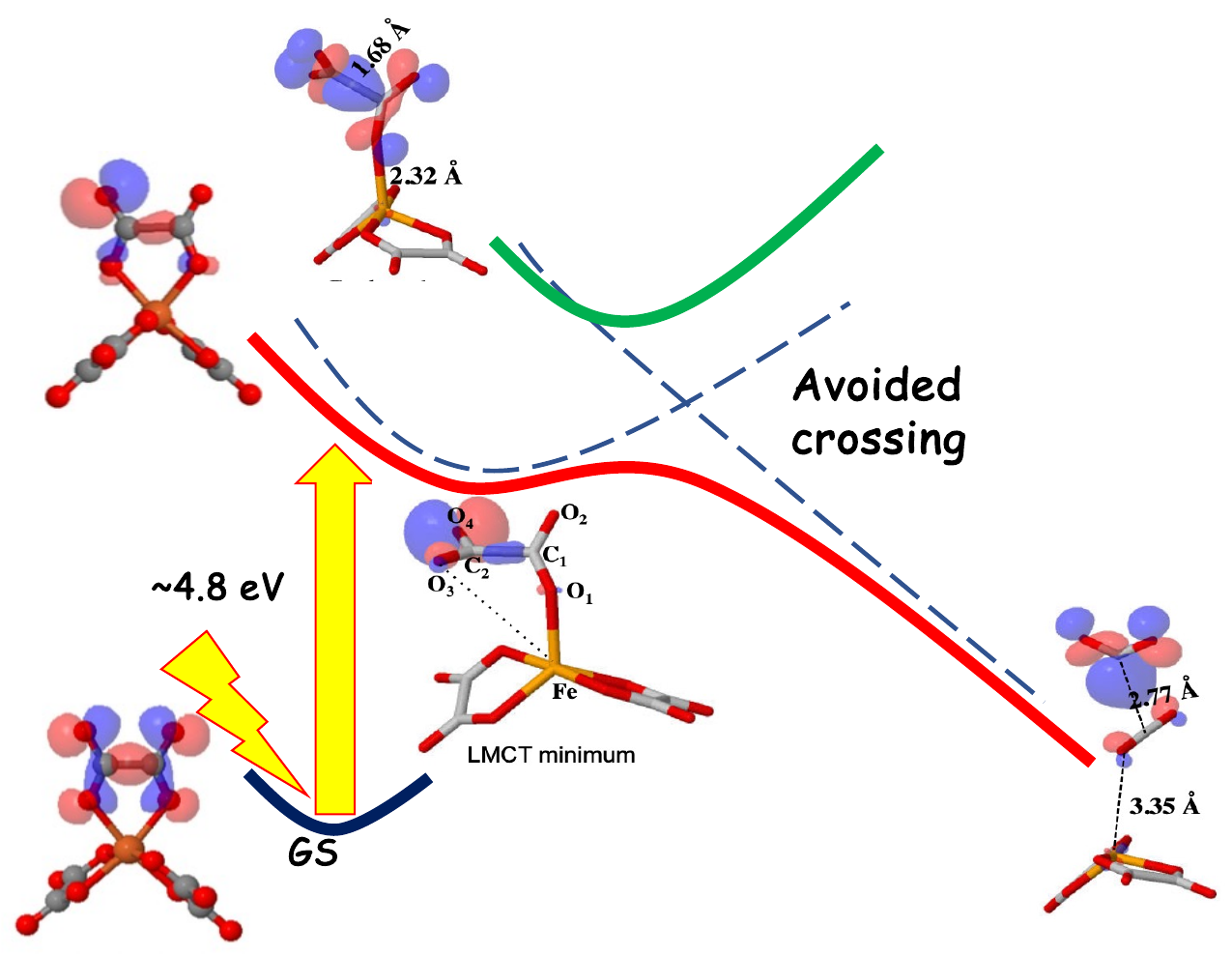}
  \caption{Primary processes in the ferrioxalate photolysis predicted using $W\Gamma$-CASSCF method.
  The singly-occupied oxalate orbital is shown in each image, which helps clarify the
  difference between the b-LMCT and d-LMCT characters on either side
  of the avoided crossing.}
  \label{fig:overall}
\end{figure}

In general, our theoretical results for ferrioxalate
are consistent with the prompt dissociation
the picture suggested by multiple experimental studies.
Based on our findings, a likely photolysis pathway is summarized
in Figure~\ref{fig:overall}.
We see that two LMCT states contribute to the photolysis of ferrioxalate:
the bound b-LMCT and the dissociative d-LMCT states,
with the low-energy pathway from the bound excited state minimum
traversing a low-barrier avoided crossing in which the adiabatic LMCT
state changes character between b-LMCT and d-LMCT.
Our results also indicate that although one Fe--O bond is likely
to rupture before the C--C bond, the other Fe--O bond then follows the
C--C bond in the latter stage of dissociation.
After the photoexcitation of $\sim$4.8 eV, the b-LMCT state is
responsible for the first Fe--O bond dissociation.
It has a bound geometry minimum at $d_{\text{FeO1}} \simeq 2.2$\r{A}
and $d_{\text{FeO3}}\simeq 4.1$\r{A}.
This finding helps clarify conflicting reports about whether a Fe--O
bond or the C--C bond cleaves first during photoexcitation.
The detachment of CO$_2$ can then proceed via an avoided crossing
with a barrier much lower than the photoexcitation energy,
through which the b-LMCT state converts in character to the d-LMCT
state, upon which the C--C $\sigma-$bond and the second Fe--O
bond then dissociate in an essentially concerted fashion with the C--C
bond leading the way.

\section*{Method}
In our state-specific approach (SS-CASSCF), we retain the two-step optimization approach in which the orbitals and CI variables are optimized separately. The CI variables are chosen at each macro iteration as the CASCI root most similar to the desired state by measuring the criterion \cite{tran2019tracking}
\begin{equation}
\label{eqn:qwg}
Q_{W\Gamma} = \frac{\braket{\Psi|(\omega-\hat{H})^2|\Psi}}{\braket{\Psi|\Psi}} +
 \frac{\hspace{1mm} || \hspace{.5mm} 
       \Gamma_t - \Gamma
       \hspace{.5mm} || \hspace{1mm}}
       {n_{{}_{\mathrm{CAS}}}}.
\end{equation}
Here $\omega$ is a target energy that is dynamically updated at each iteration, $\Gamma$ is our state's one-body density matrix, $\Gamma_t$ is a target density matrix (taken either from an initial CASCI or the previous iteration), and $n_{CAS}$ is the number of active orbitals.
The idea is that, after finding the spectrum of CASCI roots, we select the root that minimizes $Q_{W\Gamma}$
and then perform an orbital optimization step to make that root's energy stationary with respect to orbital changes. Note that the final energy depends only on the wave function at the stationary energy point, independent of $\omega$ and $\Gamma_t$.

Once the energy is stationary, there is no need for response calculations, as full energy stationarity eliminates the wavefunction response terms from the gradient expression. With our analytic gradients in hand, we perform our geometry relaxations using the {\it geomopt} module within pySCF, \cite{pyscf2018}
which interfaces with geomeTRIC \cite{wang2016geometry} and PyBerny \cite{hermann2020pyberny} for constrained and unconstrained geometry optimizations, respectively.
SA-CASSCF calculations were performed using PySCF and Molpro \cite{werner2012molpro}, and TD-DFT calculations were performed using Orca\cite{neese2022software}. The def2-SVP basis set \cite{weigend2005balanced} and the spin state sextet ($S=5/2$) were used for all calculations.

\section*{Acknowledgements}

This work was supported by the Early Career Research Program of the Office of Science, Office of Basic Energy Sciences, the U.S. Department of Energy, Grant No. DE-SC0017869. 
Calculations were performed using the Berkeley Research Computing Savio cluster.

\section*{Supporting information}
An overview of recent experimental works. Orbitals selected in CAS(7e,6o) and CAS(11e,8o). Cartesian coordinates of ground state, b-LMCT minimum optimized by SS-CASSCF and TD-DFT, stretched CC (2.2 \r{A}) equilibrium and TS geometries. 

\bibliography{main}
\end{document}